\documentclass[a4paper,11pt]{article}

\usepackage{jcappub}

\usepackage{slashed}
\usepackage{youngtab}

\newcommand\fverb{\setbox\fverbbox=\hbox\bgroup\verb}
\newcommand\fverbdo{\egroup\medskip\noindent%
            \fbox{\unhbox\fverbbox}\ }
\newcommand\fverbit{\egroup\item[\fbox{\unhbox\fverbbox}]}
\newbox\fverbbox
\newcommand{\nablaslash}{\not{\hbox{\kern-3pt $\nabla$}}}
\newcommand{\nn}{\nonumber}
\newcommand{\na}{\nabla}

\title{\boldmath Modified Brans-Dicke theory with space-time anisotropic parameters}

\author[a]{Taeyoon~Moon}
\affiliation[a]{Center for Quantum Space-time\\ Sogang University\\
Seoul 121-742, Korea}
\author[b]{and~Phillial~Oh}
\affiliation[b]{Department of Physics and Institute of Basic Science\\ Sungkyunkwan University\\
Suwon 440-746, Korea}

\emailAdd{dpproject@skku.edu}
\emailAdd{ploh@skku.edu}

\abstract{We consider the ADM formalism of the Brans-Dicke theory
and propose a space-time anisotropic extension of the theory by
introducing five free parameters. We find that the resulting theory
reveals many interesting aspects which are not present in the
original BD theory. We first discuss the ghost instability and
strong coupling problems which are present in the gravity theory
without the full diffeomorphism symmetry and show that they can be
avoided  in a region of the parameter space. We also perform the
post-Newtonian approximation and show that the constraint of the
Brans-Dicke parameter $\omega_{{\rm BD}}$ being large to be
consistent with the solar system observations could be evaded in the
extended theory. We also discuss that accelerating Universe can be
achieved without the need of the potential for the Brans-Dicke
scalar.}

\begin{document}

\maketitle
\flushbottom

\section{Introduction}

There are many alternative theories
and extensions of the Einstein's general relativity.
Motivations to consider them are diverse.
They range from a classical straightforward extension of general relativity like the  scalar-tensor theory to quantum gravity such as string theory.
 A comprehensive review can be found in  recent articles
\cite{Clifton:2011jh}.
 The classical modification is based on the observational rationale that the predictions of the general relativity has  been successful only on the solar system scale and has never been tested on the cosmological scale. Recently, it has another input from cosmology, that is, to search for the theoretical foundation of the accelerating Universe \cite{perm} has become  one of the most fundamental problems in modern cosmology. There are two main avenues. The first one is to assume an unknown source of energy which is repulsive in nature referred to as
the dark energy and various proposals for its origin have been put forward \cite{nojiri1}.  The other is to consider an alternative theory of the Einstein's theory of general relativity and modify the gravity in the IR limit.
We are still awaiting more accurate observational data to distinguish
between them.
 Therefore, modification of general relativity is
 a very important subject and
is  attracting a great deal of interest with extensive research
activities being reported (see \cite{Clifton:2011jh} and references
therein).

Among the many alternatives, scalar-tensor theory \cite{fuji} is
especially interesting. In this theory, the scalar field enters  in
a nontrivial manner, specifically through non-minimal coupling term
and it is based on solid foundation of general relativity. Typical
examples would be the Brans-Dicke (BD) theory  \cite{Brans:1961sx}
and the gravity with a dilaton field arising for instance in the
string theory. The scalar field in this type of theory provides many
distinctive theoretical aspects, for example like induced gravity
\cite{Zee:1978wi} and  explanation of the behavior of the universe
in the early inflationary stage as well as the late stage
\cite{fuji} in the cosmological context.

Especially, we attempt to modify the BD theory which is known as the
prototype of the scalar-tensor theory and is one of the simplest
alternative to Einstein's theory of general relativity (GR). BD
theory was firstly designed to properly incorporate the Mach's
principle into General relativity by replacing the gravitational
constant $G$ with a scalar field $\phi$, which can vary with space
and time. It is well-known that the observational constraints on BD
theory is restricted by the astronomical tests in the solar system,
i.e., the BD parameter $\omega_{{\rm BD}}>50000$, being obtained
from the observable Post-Newtonian parameter $\gamma_{{\rm
obs}}-1=(2.1\pm2.3)\times 10^{-5}$
\cite{Bertotti:2003rm,Will:2001mx}.

Modifications of Brans-Dicke theory have been actively pursued. Most notable is the introduction of the potential for the Brans-Dicke scalar.
For example, in Ref. \cite{Bertolami:1999dp,Sen:2000zk}, BD theory with a suitable self-interacting potential for the BD scalar field was proposed to account for the accelerated expanding of the Universe. In another example,
it can be shown that $f (R)$ gravity in the metric formalism is equivalent to BD theory with a potential and the parameter $\omega_{{\rm BD}}=0$ \cite{DeFelice:2010aj}.

Another alternative
is a space-time anisotropic extension. It is motivated by the
Horava-Lifshitz (HL) gravity \cite{Horava:2009uw} which was proposed
as a quantum gravity theory.
 It is based on anisotropic scaling of
space and time as a fundamental symmetry,
 abandoning the Lorentz symmetry at short distance.
 In the IR limit, the theory reduces to GR when the  Lorentz
  violating parameter $\lambda$ becomes 1. When $\lambda\neq1$, the
  HL gravity becomes a Lorentz violating Einstein-Hilbert action and
  the extended HL gravity \cite{Blas:2009qj,Moon:2011xi} reduces to
  Lorentz violating scalar-tensor theory in IR.
There has been considerable interest in the HL gravity theory and
various aspects have been investigated \cite{Mukohyama:2010xz}.
Among others, a BD type of generalization of the Horava-Lifshitz
gravity was shown to be possible within the detailed balance
condition
 in Ref. \cite{Lee:2010iu}, and the resulting theory reduces
 in the IR limit to the usual BD theory with a negative cosmological
 constant. On the other hand, our main interest is to investigate the possibility
of modifying the original BD theory in such a way that the resulting
theory can accommodate space-time anisotropy which breaks Lorentz
symmetry at low energy and we will test whether such a theory is
phenomenologically viable. From the theoretical point of view,
breaking of the full diffeomorphism invariance coming from the
anisotropy puts strong constraint in the resulting theory.
The breaking causes the theory to have
an additional scalar degree of freedom, whose behavior could create
serious problems, such as a ghost or classical instability or the
strong coupling problem.
On the experimental side, it should pass the solar system test which
requires $\omega_{{\rm BD}}$ has to be big in the original BD
theory.

In this paper, we propose a more straightforward modification of BD theory
starting from the ADM formalism and discuss the above issues.
A close look at the ADM formalism reveals that the original BD theory can be extended by introducing  four more free parameters,
which is related with space-time anisotropy, thereby breaking Lorentz symmetry.
Depending on the values of these parameters, the theory shows many
interesting aspects which are not present in the original BD theory.
For example, the large value of $\omega_{{\rm BD}}$ in the original BD theory
to be consistent with the solar system observations could be evaded in the modified theory.
It can be also shown that the ghost instability and strong coupling problems which are present in the gravity theory without the full diffeomorphism symmetry can be avoided  within some range of the free parameters.
Another distinctive feature is that the accelerating Universe
could be achieved without the help of some potential
contrary to the original BD theory.

The paper is organized as follows. In Sec. 2, we consider space-time anisotropic BD action with five parameters which reduces to the ordinary
(isotropic) BD action for particular choices of the four parameters.
In Sec. 3, we consider a quadratic action through the
perturbation analysis of the space-time anisotropic BD gravity  in order to investigate the pathological behaviors of scalar
graviton mode,. In Sec. 4, we perform the
perturbation analysis up to cubic order and check whether the strong
coupling problem can be cured to this order. In Sec. 5 we study the
observational constraint on the space-time anisotropic BD theory and compare with
the experimental results. The conclusion and discussions are given in
Section 6.

\section{Space-time anisotropic Brans-Dicke action}

In order to construct a space-time anisotropic Brans-Dicke gravity
with additional  parameters, let us first consider an isotropic
Brans-Dicke gravity in the ADM formalism, whose metric is
parameterized by
\begin{eqnarray}\label{admm}
ds_{{\rm ADM}}^2=-N^2dt^2+g_{ij}(dx^i+N^idt)(dx^j+N^jdt),
\end{eqnarray}
where $N$ is the lapse function, $N_i$ is the shift function, and
$g_{ij}$ is the three dimensional metric. For the ADM metric,
Brans-Dicke action is given by \cite{Lee:2010iu}
\begin{eqnarray}\label{bd0}
S_{{\rm BD}}&=&\int d^4x\sqrt{-g_{(4)}}\left\{\phi
R_{(4)}-\frac{\omega_{{\rm BD}}}{\phi}
g_{(4)}^{\mu\nu}\partial_\mu\phi\partial_\nu\phi\right\}\nn\\
&=&\int dtd^3x N\sqrt{g}\Bigg\{\phi(K_{ij}K^{ij}-K^2)
-2K\pi+\omega_{{\rm BD}}\phi^{-1}\pi^2\nn\\
&&\hspace*{8em}+\phi R-2g^{ij}\nabla_i\nabla_j\phi -\omega_{{\rm
BD}}\phi^{-1}g^{ij}\nabla_i\phi\nabla_j\phi\Bigg\},
\end{eqnarray}
where $\phi$ and $\omega_{{\rm BD}}$ is the Brans-Dicke scalar and
Brans-Dicke parameter respectively. $g_{ij}$ and $R$ are the three dimensional metric tensor and
the Ricci scalar. The extrinsic curvature $K_{ij}$ and
$\pi$ take the forms
\begin{eqnarray}
&&K_{ij}= \frac{1}{2N}(\dot{g}_{ij}-\nabla_i N_j-\nabla_jN_i ),\\
&&\pi=\frac{1}{N}(\dot{\phi}-\na_{i}\phi N^{i}),
\end{eqnarray}
where the dot
 denotes differentiation
with respect to $t$. The first line of \eqref{bd0} is the kinetic
(K) term and second line is the potential (V) term.

We first notice that the BD gravity \eqref{bd0} preserves the full
diffeomorphism symmetry. We can extend it to space-time anisotropic
BD (aBD) theory by introducing free parameters which explicitly
break the diffeomorphism invariance and consider the most general
action as follows \footnote{This extension is based on the fact that
each term of $2K\pi,~2g^{ij}\nabla_i\nabla_j\phi,~ \phi^{-1}\pi^2$,
and $\phi^{-1}g^{ij}\nabla_i\phi\nabla_j\phi$ in the action
\eqref{bd0} is invariant under foliation-preserving diffeomorphism
\eqref{Diff}. We note that the action \eqref{hlbd} is clearly
different from a scalar tensor theory written in the unitary gauge
\cite{Cheung:2007st,Gubitosi:2012hu}. This is because the scalar
tensor model obtained from applying the St\"{u}ckelberg trick does
not produce our model.}
\begin{eqnarray}\label{hlbd}
S_{{\rm aBD}}=S^{{\rm K}}_{{\rm aBD}}+S^{{\rm V}}_{{\rm aBD}},
\end{eqnarray}
where
\begin{eqnarray}
&&S^{{\rm K}}_{{\rm aBD}}= \int \,dt
d^3xN\sqrt{g}~\left\{\phi\left(K_{ij}K^{ij} -\lambda
K^2\right)-2\eta_1 K\pi+\omega_1 \phi^{-1}\pi^2\right\},\label{hlbdk}\\
&&S^{{\rm V}}_{{\rm aBD}}= \int \,dt d^3xN\sqrt{g}~ \left(\phi
R-2\eta_2
\nabla_{i}\nabla^{i}\phi-\omega_2\phi^{-1}\nabla_{i}\phi\nabla^{i}\phi\right).
\label{hlbdv}
\end{eqnarray}
 In this action, the
parameters $\lambda,~\eta_{1,2},~\omega_{1,2}$ are dimensionless
constants. Note that when
$\lambda\neq1$, $\eta_1\neq1,~\eta_2\neq1$, and
$\omega_1\neq\omega_2$, the action (\ref{hlbd}) does not have the
full diffeomorphism invariance, but is invariant under the
foliation-preserving diffeomorphism:
\begin{eqnarray}
\label{Diff}
\delta x^i &=&-\zeta^i (t, {\bf x}), ~\delta t=-f(t), \nonumber \\
 \delta
g_{ij}&=&\partial_i\zeta^k g_{jk}+\partial_j \zeta^k g_{ik}+\zeta^k
\partial_k g_{ij}+f \dot g_{ij},\nonumber\\
\delta N_i &=& \partial_i \zeta^j N_j+\zeta^j \partial_j
N_i+\dot\zeta^j
g_{ij}+f \dot N_i+\dot f N_i, \nonumber \\
\delta N&=& \zeta^j \partial_j N+f \dot N+\dot f N, \nonumber\\
\delta \phi &=&\zeta^k
\partial_k \phi+f \dot \phi .
\end{eqnarray}

We point out that the parameter
$\eta_1 (~\eta_2) $ is associated with deviation from the BD theory
along the normal (spatial) direction of the leaves of the foliation.
Moreover, we introduced two BD parameters $\omega_1$ and $\omega_2$
which need not to be the same in the anisotropic case.
With the choice of the parameters
of $\lambda=1,~\eta_1=\eta_2=1$, and $\omega_1=\omega_2=\omega_{{\rm
BD}}$, the action (\ref{hlbd}) reduces  to (isotropic)
Brans-Dicke action \eqref{bd0}. For a constant scalar field,
it becomes Lorentz violating Einstein-Hilbert action with an
anisotropic parameter $\lambda$. Note also that the space-time anisotropic BD action
(\ref{hlbd}) is still invariant with respect to global (isotropic)
conformal transformation
\begin{eqnarray}\label{gconformal}
N\to\Omega N,~~~N_i\to\Omega^2 N_i,~~~g_{ij}\to\Omega^2
g_{ij},~~~\phi\to\Omega^{-2}\phi,
\end{eqnarray}
where $\Omega$ is an arbitrary constant. We further remark that for
$\Omega=\Omega({t,{\bf x}})$, the conformal symmetry under the
transformation (\ref{gconformal}) can be extended to local conformal
symmetry if one chooses
\begin{eqnarray}\label{lcs}
\eta_1=-\frac{1}{2}(1-3\lambda),
~~~\omega_1=\frac{3}{4}(1-3\lambda),
~~~\eta_2=1,~~~\omega_2=-\frac{3}{2},
\end{eqnarray}
which implies that for $\lambda=1$ case, i.e.,
$\eta_1=\eta_2=1,~\omega_1=\omega_2=-3/2$, the space-time anisotropic BD action
(\ref{hlbd}) reduces to a conformally invariant isotropic action
\cite{Moon:2010wq}.

\section{Scalar graviton mode in the quadratic action}

In order to investigate the behavior of graviton mode in the
Minkowski background ($g_{ij}=\delta_{ij},~\phi=\phi_0$), we first
consider the following scalar perturbations of the metric and scalar field for the Minkowski background up to the linear order
\begin{eqnarray}\label{pertur}
N=e^{\alpha},~~N_i=\partial_i\beta,~~g_{ij}=e^{-2\psi}\delta_{ij} ,
~~\phi=\phi_0+\varphi.
\end{eqnarray}
In these perturbations, $\alpha,~\beta,~\psi,$ and $\varphi$ are
functions of space and time. Substituting the above perturbations
into the action (\ref{hlbd}), one finds the quadratic action
($\Box\equiv
\partial_i \partial^i$)
\begin{eqnarray}\label{qaction}
&&\hspace*{-1.2em}S^{(2)}=\int \,dt d^3x\Big\{
3\phi_0(1-3\lambda)\dot{\psi}^2+2\phi_0 (1-3\lambda)
\dot{\psi}\Box\beta+\phi_0(1-\lambda)(\Box\beta)^2
+2\eta_1(3\dot{\varphi}\dot{\psi}+\dot{\varphi}\Box\beta)\nn\\
&&\hspace*{5em}+\omega_1\phi_0^{-1}\dot{\varphi}^2+4\varphi\Box\psi+4\phi_0\alpha\Box\psi
-2\phi_0\psi\Box\psi-2\eta_2\alpha\Box\varphi-\omega_2\phi_0^{-1}
(\partial_{i}\varphi)^2\Big\}.
\end{eqnarray}
Variation of the fields $\alpha$ and $\beta$ of the quadratic action
leads to the (local) Hamiltonian and momentum constraints
\begin{eqnarray}
&& \varphi-\frac{2}{\eta_2}\phi_0\psi=0, \label{phieq}\\
&&\Box\beta+\frac{1}{1-\lambda}\left(1-3\lambda
+\frac{2\eta_1}{\eta_2}\right)\dot{\psi}=0, \label{betaeq}
\end{eqnarray}
respectively.

 From the constraints (\ref{phieq}) and (\ref{betaeq}),
we can
replace the fields $\varphi$ and $\Box\beta$ with $\psi$. After
taking the integration by parts, the quadratic action
\eqref{qaction} can be rewritten as
\begin{eqnarray}\label{quadratic}
S^{(2)}=2\int \,dt d^3x\left\{-\frac{1}{c_{\psi}^2}\dot{\psi}^2
+\left(\frac{2\omega_2}{\eta_2^2}
+\frac{4}{\eta_2}-1\right)\phi_0\psi\Box\psi\right\},
\end{eqnarray}
where
\begin{eqnarray}
c_{\psi}^2=\frac{1-\lambda}{\phi_0}\left\{3\lambda-1
+\frac{2\eta_1^2}{\eta_2^2} -\frac{4\eta_1}{\eta_2}
+\frac{2(\lambda-1)\omega_1}{\eta_2^2}\right\}^{-1}\label{condi}.
\end{eqnarray}
 We notice two things. First is that the perturbation $\varphi$ can be completely
 eliminated due to the Hamiltonian constraint (\ref{phieq}) in favor of the scalar graviton mode. The other is
 that  in the limit\footnote{This limit of $\eta_2\to\infty$
 is reminiscent of
 $\alpha\to\infty$ which is the coefficient of $(\nabla_iN/N)^2$ term introduced
 in Ref.\cite{Blas:2009qj} (see also Ref.\cite{Sotiriou:2010wn} for details).} $\eta_2 \rightarrow \infty $, we find  the action
 is exactly the same with the quadratic action in the HL gravity (for $\phi_0=1$)
\begin{eqnarray}
S^{(2)}_{{\rm HL}}=2 
\int \,dt d^3x\left\{-\frac{1}{c^2_{{\rm HL}}} \dot{\psi}^2-
\psi\Box\psi\right\},\label{HL2}
\end{eqnarray}
with $c^2_{{\rm HL}}=({1-\lambda})/({3\lambda-1})$. It can be easily
checked that in the quadratic action \eqref{HL2}, the scalar
graviton\footnote{It is well-known that a propagating scalar
graviton mode does not exist in Einstein-Hilbert action with the
full diffeomorphism invariance. More precisely, one can check from
Eq.(\ref{qaction}) that in Einstein-Hilbert action ($\lambda=1,
\varphi=0$, $\eta_{1,2}=\omega_{1,2}=0$) and conformally equivalent
case ($\lambda=1$,~$\eta_{1,2}=1,~\omega_{1,2}=-3/2$), the
Hamiltonian and momentum constraints yield $S_{\psi}^{(2)}=0$, which
implies that there is no the propagating scalar graviton mode (see
Appendix for details).} $\psi$ has serious problems, i.e., classical
instability in the case of $c^2_{{\rm HL}} <0$ or the presence of
ghost when $c^2_{{\rm HL}}>0$. Interestingly, in our framework such
pathological behaviors do not show up with
\begin{eqnarray}
c^2_{\psi}<0 ,~~0<\frac{2\omega_2}{\eta_2^2}
+\frac{4}{\eta_2}-1,  \label{condition1}
\end{eqnarray}
which can be satisfied with a wide range of the parameters region.

Before passing, we make the following remarks on the number of
scalar degrees of freedom in our proposed model. There seem to be
two scalar modes that are propagating at the perturbation level; one
coming from the BD scalar and the other scalar from the
gravitational sector which is present whenever $\lambda\neq1$
\cite{Horava:2009uw}. However, as we see from the perturbation
equation \eqref{phieq} the Hamiltonian constraint yields a direct
connection between the two scalars. It is to be noticed that this
kind of relation is not peculiar to space-time anisotropic theory.
It is persistent even in the covariant theory as long as the scalar
field is non minimally coupled like ``$\phi R$'', and {\it this
coupling is the origin of the connection between the two scalar
modes}. That is why there is only one scalar degree of freedom in
the pure BD theory, for example, which corresponds to $\eta_{2}=1$
in Eq.\eqref{phieq}. In fact, this would be true to more generalized
non minimal coupling like the scalar tensor theory of the type
$f(\phi)R$. However, it can be checked\footnote{We would like to
thank the anonymous referee for suggesting this point.} that on a
cosmological background with time dependent $\phi_0(t)$,
$\alpha^{2}$ term is present in the quadratic action and
subsequently, one is left with   two independent degrees of freedom
$\psi$ and $\varphi$. In the Appendix we elaborate on these points
by explicitly showing the details.

 Let us focus on the special cases    $\eta_1=\eta_2$ (otherwise noticed)
 which have rather special consequences. For  $\eta_1=\eta_2\equiv \eta$, the quadratic action  (\ref{quadratic}) becomes
\begin{eqnarray}\label{quadratic1}
S^{(2)}=2\phi_0\int \,dt
d^3x\left\{\left(3+\frac{2\omega_1}{\eta^2}\right)\dot{\psi}^2
+\left(\frac{2\omega_2}{\eta^2}
+\frac{4}{\eta}-1\right)\psi\Box\psi\right\},
\end{eqnarray}
and it shows that the dependency on the parameter $\lambda$ for the scalar graviton completely disappears.
 This also implies the quadratic action (\ref{quadratic}) with (\ref{condi})
 is devoid of the singularity associated
 with the limit $\lambda\to1$. The condition (\ref{condition1}) of avoiding the classical instability
 and ghost mode in this case is given by
\begin{eqnarray}
\omega_1> -\frac{3}{2}\eta^2, ~~~\omega_2>-2+\frac{1}{2}(\eta-2)^2.
\label{no-ghost}
\end{eqnarray}
Note that $\omega_2$ always has to be greater than $-2$ for any
value of $\eta$. For $\eta=1$,  $\omega_1>-\frac{3}{2}$ and
$\omega_2>-\frac{3}{2}$, which coincides the ghost free case of the
isotropic BD theory \cite{Banerjee:2000mj}, $\eta_1=\eta_2=1,
~\omega_1=\omega_2=\omega_{{\rm BD}}>-\frac{3}{2}$. Another case of
interest is when $\eta=2$. In this case, we have $\omega_1>-6,~
\omega_2>-2$. We will see that for this particular value, $\omega_2$
does not have to be a large value in order to pass the solar test as
in the original BD theory [see Eq. (\ref{gppn})]. As $\eta$ becomes
larger, $\omega_1$ can be more negative, whereas $\omega_2$ shifts
away from the value $-2$ in the positive direction.


\section{Strong coupling in the cubic action}

In the previous section, we investigated the behavior of graviton
mode in the quadratic action and showed that the pathological behavior
can be cured in a wide range of parameters with  the condition of
Eq. (\ref{condition1}) being satisfied.
Another important task is to examine if the theory gets strongly
coupled at the cubic action level in the limit of $\lambda\to1$
\cite{Koyama:2009hc,Papazoglou:2009fj}.

In order to study with the cubic order interaction terms \cite{Mald:0210},
 we first recall the non-linear scalar perturbations \eqref{pertur}
around the Minkowski background and substitute this into the action
\eqref{hlbd}.
 After some tedious manipulations
 one can find that the cubic-order action is given by
\begin{eqnarray}
&&S^{(3)}~=~\nn\\
&&\int \,dt d^3x\Bigg\{-2\varphi(\partial\psi)^2
+2\phi_0\psi(\partial\psi)^2-2\alpha\phi_0(\partial\psi)^2-4\varphi\psi\Box\psi
+4\alpha\varphi\Box\psi+2\phi_0\alpha^2\Box\psi\nn
\\&&+2\phi_0\psi^2\Box\psi
-4\phi_0\alpha\psi\Box\psi-\eta_2\alpha^2\Box\varphi-\eta_2\psi^2\Box\varphi
+2\eta_2\alpha\psi\Box\varphi-2\eta_2\psi\partial_i\varphi\partial_i\psi
\nn\\&&+2\eta_2\alpha\partial_i\varphi\partial_i\psi
+\omega_2\phi_0^{-1}\psi(\partial\varphi)^2
-\omega_2\phi_0^{-1}\alpha(\partial\varphi)^2+\omega_2\phi_0^{-2}\varphi(\partial\varphi)^2
-9(1-3\lambda)\phi_0\psi\dot{\psi}^2\nn\\
&&-2(1-3\lambda)\phi_0\psi\dot{\psi}\Box\beta
-2(1-3\lambda)\phi_0\dot{\psi}\partial_{i}\psi\partial_{i}\beta
-2(1-\lambda)\phi_0\Box\beta\partial_i\psi\partial_i\beta
+\phi_0\psi(\partial_i\partial_j\beta)^2
\nn\\
&&+4\phi_0\partial_i\partial_j\beta\partial_i\beta\partial_{j}\psi
-\lambda\phi_0\psi(\Box\beta)^2-3(1-3\lambda)\phi_0\alpha\dot{\psi}^2
-2(1-3\lambda)\phi_0\alpha\dot{\psi}\Box\beta
\nn\\
&&-\alpha\phi_0(\partial_i\partial_j\beta)^2+\phi_0\lambda\alpha(\Box\beta)^2
+3(1-3\lambda)\varphi\dot{\psi}^2+2(1-3\lambda)\varphi\dot{\psi}\Box\beta
+\varphi(\partial_i\partial_j\beta)^2\nn\\&&-\lambda\varphi(\Box\beta)^2
-2\eta_1\Big(9\psi\dot{\psi}\dot{\varphi}+3\alpha\dot{\psi}\dot{\varphi}
+3\dot{\psi}\partial_i\varphi\partial_i\beta
+\dot{\varphi}\partial_i\beta\partial_i\psi+\alpha\dot{\varphi}\Box\beta
+\psi\dot{\varphi}\Box\beta\nn\\
&&+\partial_i\varphi\partial_i\beta\Box\beta\Big)
-\omega_1\Big(\phi_0^{-1}\alpha\dot{\varphi}^2+3\phi_0^{-1}\psi\dot{\varphi}^2
+\phi_0^{-2}\varphi\dot{\varphi}^2+2\phi_0^{-1}\dot{\varphi}\partial_i\varphi\partial_i\beta\Big)
\Bigg\}. \label{cubic}
\end{eqnarray}
Using the first-order Hamiltonian and momentum constraints
(\ref{phieq}), (\ref{betaeq}) obtained in the previous section, the
above action (\ref{cubic}) reduces to
\begin{eqnarray}
S^{(3)}=
 \phi_0\int \,dt d^3x
\left\{A_1
 {\psi}(\partial_i{\psi})^2
+A_2
\dot{\psi}\partial_i{\psi}\partial_i\left(\frac{\dot{\psi}}{\Box}\right)
+A_3
\psi\left(\frac{\partial_i\partial_j}{\Box}\dot{\psi}\right)^2+A_4
{\psi}\dot{\psi}^2\right\}.\label{cubic1}
\end{eqnarray}
where
\begin{eqnarray}
&&A_1=\frac{8\omega_2}{\eta_2^3}+\frac{4\omega_2}{\eta_2^2}+\frac{12}{\eta_2}-2,\nn\\
&&A_2=-\frac{8\eta_1+4(1-3\lambda)\eta_2}{(-1+\lambda)^2\eta_2^3}
\Big\{2\eta_1^2-4\eta_1\eta_2+(3\lambda-1)\eta_2^2+2(\lambda-1)\omega_1\Big\},\nn\\
&&A_3=-\frac{3\eta_2-2}{(-1+\lambda)^2\eta_2^3}\Big\{2\eta_1+(1-3\lambda)\eta_2\Big\}^2,\nn\\
&&A_4=\frac{1}{(-1+\lambda)^2\eta_2^3}\Big\{(1+\lambda)(3\lambda-1)(3\eta_2-2)\eta_2^2
-4\eta_1^2\Big(2\lambda+3\eta_2(\lambda-2)\Big)\nn\\
&&\hspace*{8em}-4\eta_1\eta_2\Big(2-6\lambda+3\eta_2(1+\lambda)\Big)
-4\omega_1(\lambda-1)^2(3\eta_2+2)\Big\}\nn.
\end{eqnarray}

Note that when 
$\eta_2\rightarrow\infty$ the above action (\ref{cubic1}) once again
reduces to the cubic action in the HL gravity as
\begin{eqnarray}
&&\hspace*{-1.3em} S^{(3)}_{{\rm HL}}= 2
\int \,dt d^3x \Bigg\{-
 {\psi}(\partial_i{\psi})^2
+\frac{2}{c^4_{{\rm HL}}}\dot{\psi}\partial_i{\psi}\partial_i\left(\frac{\dot{\psi}}{\Box}\right)\nn\\
&&\hspace*{10em}+\frac{3}{2}\left[-\frac{1}{c^4_{{\rm
HL}}}\psi\left(\frac{\partial_i\partial_j}{\Box}\dot{\psi}\right)^2
+\frac{2c^2_{{\rm HL}}+1}{c^4_{{\rm
HL}}}{\psi}\dot{\psi}^2\right]\Bigg\}.
\end{eqnarray}
It is well-known that the above action is confronted with the strong coupling problem
in the limit $\lambda\rightarrow 1$ \cite{Koyama:2009hc,Papazoglou:2009fj}. On the other hand, in the case of
$\eta_1=\eta_2\equiv \eta$ of our main focus, the above coefficients $A_2,~A_3,~A_4$ considerably simplify to become
\begin{eqnarray}
&&A_2=12\left(3+\frac{2\omega_1}{\eta^2}\right),
~~~A_3=-9\left(3-\frac{2}{\eta}\right),
~~~A_4=9-\frac{6}{\eta}-\frac{4\omega_1(2+3\eta)}{\eta^3}\nn,
\end{eqnarray}
which again shows that there is no dependency on the parameter $\lambda$ for the
scalar graviton up to the cubic order.
To see this more closely, we
introduce the canonical variable
$\hat{\psi}=\sqrt{2}\psi/|c_{\psi}|$ with
$|c_{\psi}|^{-2}=\phi_0(3+2\omega_1/\eta^2)$, in terms of which the
quadratic action (\ref{quadratic1}) becomes
\begin{eqnarray}
S^{(2)}=\int \,dt d^3x\left\{\dot{\hat{\psi}}^2
+\left(\frac{2\omega_2}{\eta^2}
+\frac{4}{\eta}-1\right)\phi_0|c_{\psi}|^2\hat{\psi}\Box\hat{\psi}\right\}.
\end{eqnarray}
By using the canonical variable $\hat{\psi}$, the cubic action
(\ref{cubic1}) can be written as
\begin{eqnarray}
&&\hspace*{-2em}S^{(3)}=
 \frac{1}{2\sqrt{2}}\int \,dt d^3x
\left\{\left(\frac{8\omega_2}{\eta^3}+
\frac{4\omega_2}{\eta^2}+\frac{12}{\eta}-2\right)
\phi_0|c_{\psi}|^3
 {\hat\psi}(\partial_i{\hat\psi})^2
+12|c_{\psi}| \dot{\hat\psi}\partial_i{\hat\psi}\partial_i
\left(\frac{\dot{\hat\psi}}{\Box}\right)\right.\nonumber\\
 &&\hspace*{-3em}\left.-9\left(3-\frac{2}{\eta}\right)\phi_0|c_{\psi}|^3
\hat\psi\left(\frac{\partial_i\partial_j}{\Box}\dot{\hat\psi}
\right)^2+\left(\left(27+\frac{6}{\eta}\right)
\phi_0|c_{\psi}|^3-2\left(3+\frac{2}{\eta}\right)|c_{\psi}|\right)
{\hat\psi}\dot{\hat\psi}^2\right\}.\label{cubic2}
\end{eqnarray}
This action clearly shows non dependency on the parameter $\lambda $ and that  we do not need to consider any
fine-tuning for the interactions to be regular.

 Before
closing this section, we comment on the case with local conformal invariance. It turns out that with the choice of the parameters as in (\ref{lcs}),
the time dependent cubic terms of $\psi$ in the action
(\ref{cubic1}) vanish and  we get
\begin{eqnarray}
S^{(3)}=
 \phi_0\int \,dt d^3x
\left\{-8
 {\psi}(\partial_i{\psi})^2
\right\}\label{cubic3}.
\end{eqnarray}
This shows  that  the strong coupling problem does
not show up at the cubic-order perturbation of the action
with the local conformal invariance.

\section{Cosmological constraint on space-time anisotropic BD theory}

In this section we investigate the cosmological tests, which can
provide the observational constraints for alternative theories of
gravitation. For this purpose, we first consider the solar system
test. Taking into account the static part of the perturbations in
Sec.3, the Lagrangian (\ref{qaction}) with a static point-like
source term of mass $M_s$ can be written by
\cite{Steinhardt:1994vs,Olmo:2005zr}
\begin{eqnarray}
{\cal L}_{{\rm static}}=\Big\{4\varphi\Box\psi+4\phi_0\alpha\Box\psi
-2\phi_0\psi\Box\psi-2\eta_2\alpha\Box\varphi-\omega_2\phi_0^{-1}
(\partial_{i}\varphi)^2\Big\}-M_s\alpha\delta^3({\bf x}).
\end{eqnarray}
From varying for the fields $\psi,~\alpha,$ and $\varphi$, we obtain
the following equations:
\begin{eqnarray}
\Box\varphi+\phi_0\Box\alpha-\phi_0\Box\psi&=&0,\nn\\
&&\nn\\
4\phi_0\Box\psi-2\eta_2\Box\varphi-M_s\delta^3({\bf x})&=&0,\nn\\
&&\nn\\
2\Box\psi-\eta_2\Box\alpha+\omega_2\phi_0^{-1}\Box\varphi&=&0.\label{spa}
\end{eqnarray}
The corresponding solutions to Eq. (\ref{spa}) are
\begin{eqnarray}\label{solp}
\psi=\frac{\eta_2+\omega_2}{2+\omega_2}\alpha,
~~~~~\varphi=\frac{(\eta_2-2)\phi_0}{\omega_2+2}\alpha,
\end{eqnarray}
where $\alpha$ is
\begin{eqnarray}\label{sola}
\alpha=-\frac{(\omega_2+2)M_s}{8\pi\phi_0(2\omega_2-\eta_2^2+4\eta_2)|{\bf
x}|}.
\end{eqnarray}
It is well-known that the linear expansion of the metric
(\ref{pertur}) can be expressed by the Newtonian potential $u$ and
the Post-Newtonian parameter $\gamma$ as follows:
\begin{eqnarray}
g_{00}&=&-N^2=-1-2\alpha=-1+2u,\\
g_{ij}&=&(1-2\psi)\delta_{ij}=(1+2\gamma u)\delta_{ij},
\end{eqnarray}
which implies that for the solution (\ref{solp}) $u=-\alpha$ and the parameter
$\gamma$ can be written by
\begin{eqnarray}\label{gppn}
\gamma=\frac{\psi}{\alpha}=\frac{\eta_2+\omega_2}{2+\omega_2}
=1+\frac{\eta_2-2}{\omega_2+2}.
\end{eqnarray}

Let us concentrate on the case $\eta_1=\eta_2=\eta$ again. The solar system tests currently constrain
$\gamma$ as \cite{Bertotti:2003rm}
\begin{eqnarray}\label{gbd}
|\gamma-1|<2\times 10^{-5}.
\end{eqnarray}
For the case of the BD theory ($\eta=1,~\omega_2=\omega_{{\rm
BD}}$), we see that
 the above condition ($\ref{gbd}$) together with ($\ref{gppn}$)
 restricts the region of $\omega_{{\rm BD}}$ to
\begin{eqnarray}\label{wbd}
\omega_{{\rm BD}}>50000.
\end{eqnarray}
Note, however, that unlike the BD theory, this  large value
constraint on $\omega_{{\rm BD}}$ (\ref{wbd}) can be circumvented in
the space-time anisotropic BD case when $\eta$ is very close to 2 as
can be seen from
\begin{eqnarray}\label{e2}
\left|\eta-2\right|<2\times 10^{-5}(\omega_2+2)
\end{eqnarray}
which is obtained from Eq. (\ref{gppn}). Especially, when $\eta=2,$ any $\omega_2>-2$
is allowed which is also consistent with the stability and no-ghost condition
of Eq. (\ref{no-ghost}).
In addition, for
 the solution (\ref{sola}) of $\alpha$ one can find
the effective Newton's constant $G_{eff}$ as
\begin{eqnarray}
G_{eff}=\frac{\omega_2+2}{8\pi\phi_0(2\omega_2-\eta^2+4\eta_2)},
\end{eqnarray}
being obtained from the Newtonian potential, i.e., $u=G_{eff}
M_s/|{\bf x}|$. In the General Relativity limit
($\omega_2=\omega_{{\rm BD}}\to \infty$) we have $G_{eff}\to G_{N}$
for the substitution of $\phi_0\leftrightarrow 1/(16\pi G_{N})$ with
the Newton's constant $G_{N}$. It is interesting to notice that in
the space-time anisotropic BD theory the reduction $G_{eff}\to
G_{N}$ can be achieved alternatively with appropriate choices of the
free parameters different from the BD theory. Especially, when
choosing the $\eta=2$ once again, $G_{eff}$ reduces to the Newton's
constant $G_{N}$ irrespective of the value of $\omega_2$.

On the other hand, we can also obtain the cosmological constraint
from a ratio factor of $G_{c}/G_{eff}$ ($G_{c}$ being the effective
cosmological gravitational constant defined in Eq.
(\ref{cosmological})) which is related to the primordial helium
abundance \cite{Carroll:2004ai}.  Substituting the FRW
metric\footnote{In synchronous time $t$, the cosmological ADM metric
(\ref{admm}) is given by $N=1,~N_i=0$, and $g_{ij}=a(t)^2\eta_{ij}$,
where $a(t)$ is the scale factor
\cite{Blas:2009qj,Calcagni:2009ar,Lu:2009em}, which corresponds to
the FRW metric.} ansatz and $\phi=\phi(t)$ into the action
(\ref{hlbd}) with matter term included and varying the action with
respect to the lapse function $N(t)$, we find the standard Friedmann
equation can be written as
\begin{eqnarray}
H^2=\frac{8\pi G_{c}}{3}\rho,\label{cosmological}
\end{eqnarray}
where $H$ is the Hubble parameter, $\rho$ is total matter density of the
Universe, and the effective cosmological gravitational constant $G_{c}$ is
given by
\begin{eqnarray}
G_{c}=\frac{1}{8\pi\phi(3\lambda-1)}.
\end{eqnarray}
It is pointed out that $G_c$ is equivalent to $G_N$ for the
substitution of $\phi=\phi_0\leftrightarrow 1/(16\pi G_N)$ in the
limit of $\lambda\to 1$. In our case for $\phi=\phi_0$, we obtain
the cosmological constraint from the observational bound
\cite{Carroll:2004ai} of $G_c/G_{eff}$ as
\begin{eqnarray}\label{gcgeff}
\Big|\frac{G_{c}}{G_{eff}}-1\Big|
=\Big|\frac{3(\lambda-1)}{3\lambda-1}
+\frac{(\eta-2)^2}{(3\lambda-1)(\omega_2+2)}\Big|<0.125.
\end{eqnarray}
Comparing (\ref{e2}) with the bound (\ref{gcgeff}) by using
(\ref{no-ghost}), we find that the allowed range of the parameter
$\lambda$ is
\begin{eqnarray}\label{lambdac}
-\frac{2}{3}<\lambda-1< 0.095 \approx 10^{-1},
\end{eqnarray}
which imposes a rather loose constraint around $\lambda=1$ \cite{Blas:2009qj}.

\section{Conclusion and discussion}

In this paper, we constructed a space-time anisotropic Brans-Dicke
gravity, which includes five free parameters, i.e.,
$\lambda,~\eta_{1},~\eta_2,~\omega_{1},$ and $\omega_2$. In the case
of $\lambda=\eta_{1}=\eta_{2}=1$ and $\omega_1=\omega_2=\omega_{{\rm
BD}}$, the gravity reduces to the ordinary Brans-Dicke action with a
BD parameter $\omega_{{\rm BD}}$. When fixing the scalar field
$\phi$ to be a constant value, it becomes  Lorentz-violating
Einstein-Hilbert action with an anisotropic parameter $\lambda$.

We found that in the perturbation around the Minkowski
background and constant scalar field the scalar graviton at
the quadratic as well as cubic order in the space-time anisotropic BD action does not
show any pathological behaviors within some parameter range.
This suggests that the space-time anisotropic BD gravity can be a viable theoretical model.
Especially, the case $\eta_1=\eta_2$ reveals intriguing property of $\lambda$ independence of both the quadratic and cubic actions.
 Also in the context of cosmological
model, we have checked that unlike the case of the BD theory
 which is imposed by
$\omega_{{\rm BD}}>50000$ to be consistent with the experimental observations,
this large value can be evaded in the
 space-time anisotropic BD theory with the special value of $\eta=2$ which is one of the
 novel feature of the space-time anisotropic BD theory. But the origin of these special properties associated with  $\eta=2$
 is a puzzling aspect which needs to be investigated further.

 We conclude with a couple of comments on the
issues related to the BD theory. First, it is well-known that in the
standard BD model without potential, there is no accelerated
expansion, so one has to consider a potential term
\cite{Bertolami:1999dp,Sen:2000zk}. However, it is found  in the space-time anisotropic BD
model including matter contribution that we have a de Sitter
solution ($H={\rm const}.$) for the FRW metric, given by
\begin{eqnarray}\label{de}
a=a_0 e^{Ht},~~\phi=\phi_0 e^{-3(1+\omega_m) Ht},~~\rho_m=\rho_0
a^{-3(1+\omega_m)}
\end{eqnarray}
with constants $a_0,~\phi_0,~\rho_0$ and $P_{m}=\omega_m\rho_m$. In
particular, $\lambda$ and $\rho_0$ satisfy the following relation:
\begin{eqnarray}\label{lr}
\lambda=2\eta+(1-\omega_m^2)\omega_1+1/3,~~
\rho_0=-9\omega_m\phi_0a_0^{3(1+\omega_m)}
H^2(\eta+\omega_1(1+\omega_m)).
\end{eqnarray}
For $\rho_m=0 ~({\rm vacuum})$ case, the above relation yields
\begin{eqnarray}\label{omegaf1}
\omega_1=-2\eta+\lambda-1/3.
\end{eqnarray}
It should be pointed out that in the BD limit, i.e.,
$\eta\to1,~\lambda\to1$ the parameter $\omega_1=\omega_{{\rm BD}}$
becomes a negative value, $\omega_{{\rm BD}}=-4/3$ which conflicts
with the lower bound $\omega_{{\rm BD}}> 50000$, even if it
satisfies the ghost-free condition of $\omega_{\rm BD}>-3/2$
\cite{Banerjee:2000mj}. However, in the space-time anisotropic BD case this problem can be
circumvented by choosing $\eta$ to be $\eta>4/3$ or $\eta<0$ for the
bound (\ref{lambdac}) which is obtained from substituting (\ref{omegaf1})
into the ghost-free condition of Eq. (\ref{no-ghost}). It is worth noticing that the allowed range is
not in conflict with the special case of $\eta=2$.

The second one is a speculation about quantum gravity. In
Refs.\cite{'tHooft:1974bx,Ichinose:1984cf,Cho:1989wb}, the one-loop
effective action in the pure BD theory
 is calculated and it is shown that the BD theory is not renormalizable.
However, it becomes a renormalizable theory if  curvature squared
terms  and a scale invariant self-interaction are included
\cite{Smolin:1979uz}. Likewise, we suspect that self-interacting
scale invariant anisotropic BD with curvature squared terms
\cite{Horava:2009uw,Blas:2009qj,Sotiriou:2009gy,Moon:2011xi} might
constitute a UV completion of the theory. The details are beyond the
scope of the present paper, but the subject
 deserves further investigations.
 \\
\section*{Appendix: The number of scalar modes in gravity models}

We consider the space-time anisotropic BD action
(\ref{hlbd}) and its perturbative action (\ref{qaction}), which produce various limits:\\
\\
(i) GR ($\phi=\phi_0,~\lambda=1,~\eta_{1,2}=0,~\omega_{1,2}=0$)\\
\\
For this case, the quadratic equation (\ref{qaction}) reduces simply
to
\begin{eqnarray}
&&\hspace*{-1.2em}S^{(2)}_{\rm GR}=2\phi_0\int \,dt
d^3x\Big\{-3\dot{\psi}^2-2
\dot{\psi}\Box\beta+2\alpha\Box\psi-\psi\Box\psi\Big\},\nonumber
\end{eqnarray}
which leads to the Hamiltonian and momentum constraint, $\Box\psi=0$
and $\dot{\psi}=0$, respectively. Therefore the number of scalar
modes is zero because of $S^{(2)}_{\rm GR}=0$.
\\
\\
(ii) HL gravity ($\phi=\phi_0,~\eta_{1,2}=0,~\omega_{1,2}=0$)\\
\\
The quadratic equation (\ref{qaction}) in this case reduces to
\begin{eqnarray}\label{appqaction2}
&&\hspace*{-1.2em}S^{(2)}_{\rm HL}=\Big\{\int \,dt d^3x\Big(
3\phi_0(1-3\lambda)\dot{\psi}^2+2\phi_0 (1-3\lambda)
\dot{\psi}\Box\beta+\phi_0(1-\lambda)(\Box\beta)^2
-2\phi_0\psi\Box\psi\Big)\nn\\
&&\hspace*{15em}+4\phi_0\int dt \alpha(t)\int d^3x\Box\psi\Big\}.
\end{eqnarray}
The second line in Eq.(\ref{appqaction2}) is due to the
projectability\footnote{See Refs.\cite{Blas:2009qj,Sotiriou:2010wn}
for the non-projectability case in the HL gravity.} condition of the
original HL gravity, which implies that $\alpha$ is a function of
$t$ only, $\alpha=\alpha(t)$. Variation of the fields $\alpha$ and
$\beta$ of the quadratic action leads to the Hamiltonian and
momentum constraints
\begin{eqnarray}
\int d^3x \Box\psi=0~~~~~{\rm
and}~~~~~\Box\beta+\frac{1-3\lambda}{1-\lambda}\dot{\psi}=0.
\nonumber
\end{eqnarray}
 Finally the quadratic action
becomes
\begin{eqnarray}
S^{(2)}_{{\rm HL}}=2\phi_0 
\int \,dt d^3x\left\{-\frac{1}{c^2_{{\rm HL}}} \dot{\psi}^2-
\psi\Box\psi\right\},\nonumber
\end{eqnarray}
with $c^2_{{\rm HL}}=({1-\lambda})/({3\lambda-1})$, which is the
same as (\ref{HL2}). Thus in the HL gravity the number of scalar
modes is one.
\\
\\
(iii) BD gravity ($\lambda=1,~\eta_{1,2}=1,~\omega_{1,2}=\omega_{\rm BD}$)\\
\\
In this case, the quadratic equation (\ref{qaction}) is given by
\begin{eqnarray}
&&\hspace*{-1.2em}S^{(2)}_{\rm BD}=\int \,dt d^3x\Big\{
-6\phi_0\dot{\psi}^2-4\phi_0\dot{\psi}\Box\beta
+2(3\dot{\varphi}\dot{\psi}+\dot{\varphi}\Box\beta)+\omega_{\rm
BD}\phi_0^{-1}\dot{\varphi}^2\nn\\
&&\hspace*{8em}+4\varphi\Box\psi+4\phi_0\alpha\Box\psi
-2\phi_0\psi\Box\psi-2\alpha\Box\varphi-\omega_{\rm BD}\phi_0^{-1}
(\partial_{i}\varphi)^2\Big\}\nonumber.
\end{eqnarray}
The corresponding Hamiltonian and momentum constraints are
\begin{eqnarray}
&& \Box\varphi-2\phi_0\Box\psi=0 \nonumber\\
{\rm and}~&&2\phi_0\dot{\psi}-\dot{\varphi}=0, \nonumber
\end{eqnarray}
respectively. As a result, the quadratic action can be written as
\begin{eqnarray}
S^{(2)}_{{\rm BD}}=2\phi_0 \int \,dt d^3x\left\{(3+2\omega_{\rm BD})
\dot{\psi}^2+ (3+2\omega_{\rm BD})\psi\Box\psi\right\}\nonumber.
\end{eqnarray} Therefore the number of
 scalar modes is one. Furthermore, this aspect remains even for scalar tensor gravity of the type $f(\phi) R$ with $f$
an arbitrary function of $\phi$, where the Hamiltonian constraint
would yield $\Box\varphi-2f(\phi_0)\Box\psi/f^{\prime}(\phi_0)=0$.
In this case, also a scalar and gravitational scalar modes are
interrelated and the propagating degree of freedom for scalar modes
is one.
\\

However, the above property for the flat Minkowski background does not persist on a FRW universe. This is simply because in time dependent background, the presence of $\alpha^2$ term prevents a direct connection between the two scalar modes as in Eq. (\ref{phieq}). A straightforward calculation in the FRW background\footnote{The scalar perturbations of the metric and scalar field for the FRW background are given by
\begin{eqnarray}
N=e^{\alpha},~~N_i=a^2\partial_i\beta,~~g_{ij}=a^2e^{-2\psi}\delta_{ij}
, ~~\phi=\phi_0+\varphi,\nonumber
\end{eqnarray}
where the scale factor $a$ and $\phi_0$ are function of $t$.} yields (terms containing only $\alpha$)

\begin{eqnarray}
S^{(2)}_{\alpha}&=&a^3 \int \,dt d^3x\Bigg\{\alpha\Big[\{6\phi_0H(1-3\lambda)-6\eta_1\dot{\phi}_0\}\dot{\psi}
+\{6\eta_1H-2\omega_1\phi_0^{-1}\dot{\phi}_0\}\dot{\varphi}
\nonumber\\&&+~\{9\phi_0H^2(1-3\lambda)-18\eta_1H\dot{\phi}_0+3\omega_1\phi_0^{-1}\dot{\phi}_0^{2}\}\psi
-\{3H^2(1-3\lambda)-\omega_1\phi_0^{-2}\dot{\phi}_0^{2}\}\varphi\nonumber\\&&
+~\{2\phi_0H(1-3\lambda)-2\eta_1\dot{\phi}_0\}\Box\beta+4a^{-2}\phi_0\Box\psi-2a^{-2}\eta_2\Box\varphi
\Big]\nonumber\\
&&\hspace*{4.2em}+~\alpha^2\Big[\frac{3}{2}\phi_0H^2(1-3\lambda)-3\eta_1H\dot{\phi}_0
+\frac{\omega_1}{2}\phi_0^{-1}\dot{\phi}_0^{2}\Big]\Bigg\}\label{SS},
\end{eqnarray}
which leads to the Hamiltonian constraint as follows:
\begin{eqnarray}\label{FF}
\alpha F_{1}+F_{2}(\psi,\varphi)=0.
\end{eqnarray}
Here $F_{1}$ does not contain any of the two scalar modes $\psi$,
$\varphi$ and $F_{2}(\psi,\varphi)$ is a function of the two fields.
Note that in the Minkowski limit ($a\to1$,~$H\to0$, and $\phi_0={\rm
const}.$), $F_1$  vanishes and $F_2$ yields just Eq. (\ref{phieq}).
The Hamiltonian constraint (\ref{FF}) shows that two scalar modes
are no longer dependent upon each other. Substitution of $\alpha$
from (\ref{FF}) into the full action containing all the other terms
will give a perturbed quadratic action with two independent degrees
of freedom $\psi$ and $\varphi$.

\section*{Acknowledgments}

~~~~~We would like to thank Professors Joohan Lee, Tae Hoon Lee and Mu-In Park for useful discussions.
This work was supported by the National Research Foundation of
Korea(NRF) grant funded by the Korea government(MEST) through the
Center for Quantum Spacetime(CQUeST) of Sogang University with grant
number 2005-0049409. PO was supported by the BSRP through the
National Research Foundation of Korea funded by the MEST
(2010-0021996).

\end{document}